\documentclass[%
reprint,
 amsmath,assume,
 aps,
]{revtex4-2}
\usepackage{lineno}
\usepackage{gensymb}
\usepackage{graphicx}
\usepackage{dcolumn}
\usepackage{bm}
\usepackage{siunitx}

\DeclareSIUnit\bar{bar}

\begin{document}


\title{Sub-2-Cycle, Terawatt Pulses via Double-Stage Multi-Pass Cell Compression of an Yb Laser}
\author{Victor Koltalo$^{1}$}\email{victor.koltalo@ensta.fr}
\author{Jaismeen Kaur$^{1}$}
\author{Louis Daniault$^{1}$}
\author{Antoine Cavagna$^{1}$}
\author{Micka\"el Allemand$^{3}$}
\author{Florent Pallas$^{3}$}
\author{Antoine Courjaud$^{3}$}
\author{Emilien Gontier$^{3}$}
\author{C\'edric Sire$^{2}$}
\author{François Sylla$^{2}$}
\author{Rodrigo Lopez-Martens$^{1}$}

\affiliation{$^{1}$Laboratoire d’Optique Appliquée (LOA), CNRS, Ecole polytechnique, ENSTA, Institut Polytechnique de Paris, Palaiseau, France}
\affiliation{$^{2}$SourceLAB, 7 rue de la Croix Martre, Palaiseau 91120, France}
\affiliation{$^{3}$Amplitude, 11 avenue de Canteranne, Cité de la Photonique, Bâtiment MEROPA, 33600 Pessac, France}

\date{\today}

\begin{abstract}
We report a terawatt-class, sub-2-cycle post-compressed Yb-based laser system operating at kHz repetition rate. 18 mJ, 400 fs pulses delivered at 1 kHz by a commercial Yb:YAG laser are spectrally broadened and temporally compressed in a double-stage multi-pass cell setup featuring an argon-filled Herriott-type cell followed by a helium-filled array-type cell, achieving an overall compression factor of 65. The resulting compressed pulses are measured to have 6.2 fs duration (1.8 optical cycles at 1030 nm) and 12.5 mJ energy, corresponding to 1.4 TW effective peak power. To the best of our knowledge, this constitutes the highest peak power reported to date for a few-cycle, kHz--repetition-rate, post-compressed Yb laser system. Focused intensity measurements yield a normalized vector potential of a$_0$ = 3.0, confirming the applicability of this light source for laser wakefield acceleration.
\end{abstract}

\maketitle

Few-cycle laser pulses can be used for laser-plasma acceleration (LPA) to produce stable few-MeV electron beams at kiloHertz (kHz) repetition rate for applications \cite{Faure:19}. These experiments rely on the combination of high peak power and ultrashort pulse duration, typically below 3 optical cycles, to achieve suitable acceleration conditions using few-milliJoule (mJ) pulse energies. So far, the most established route to generate such pulses has been post-compression of multi-mJ kHz Titanium:Sapphire (Ti:Sa) laser amplifiers, most commonly via self-phase modulation in hollow-core fibers \cite{Nagy:21}, which can reliably compress sub-50 fs range to few-cycle duration with excellent spatiotemporal pulse quality. However, power-scaling of kHz Ti:Sa laser systems remains complex and costly due to severe thermal management constraints, which in turn limit the deployment of post-compressed Ti:Sa lasers for LPA-based applications. To overcome these limitations and exploit the full potential of LPA for industrial applications, a strong and sustained effort has been made to temporally post-compress high average power Ytterbium (Yb)-based laser drivers in multi-pass cells \cite{Viotti2022Multipass} due to their far superior scalability in terms of average power \cite{Grebing:20, Muller:21, Pfaff2023Nonlinear}. 

Because of the relatively long native pulse duration of Yb-based lasers (typically hundreds of femtoseconds), MPC post-compression setups require two-stage \cite{Balla2020Postcompression, Muller:21, Viotti2023Fewcycle, Westerberg2026Near} or even three-stage \cite{Seres2026Over} arrangements to reach sub-10 fs pulse duration. Distributing post-compression across several MPC stages relaxes the nonlinear phase accumulated per stage, thereby ensuring good spatio-temporal quality at the large compression factors required to reach the few-cycle regime \cite{Escoto2022Temporal}. Moreover, it enables the use of dielectric mirror coatings with higher reflectivity during the early stages of spectral broadening, thereby improving the energy throughout of the overall system. Despite these advances, the simultaneous achievement of few-cycle duration, multi-mJ energy, and kHz repetition rate has remained challenging: prior Yb-based post-compression demonstrations have typically reached either sub-2-cycle duration at lower energies or higher energies at longer pulse durations, but not both together at kHz repetition rate. In this Letter, we report what is, to the best of our knowledge, the first terawatt peak-power post-compressed Yb-based laser system delivering 12.5 mJ energy 6.2 fs pulses at kHz repetition rate. 

\begin{figure*}[t!]
    \centering
    \includegraphics[width=\linewidth]{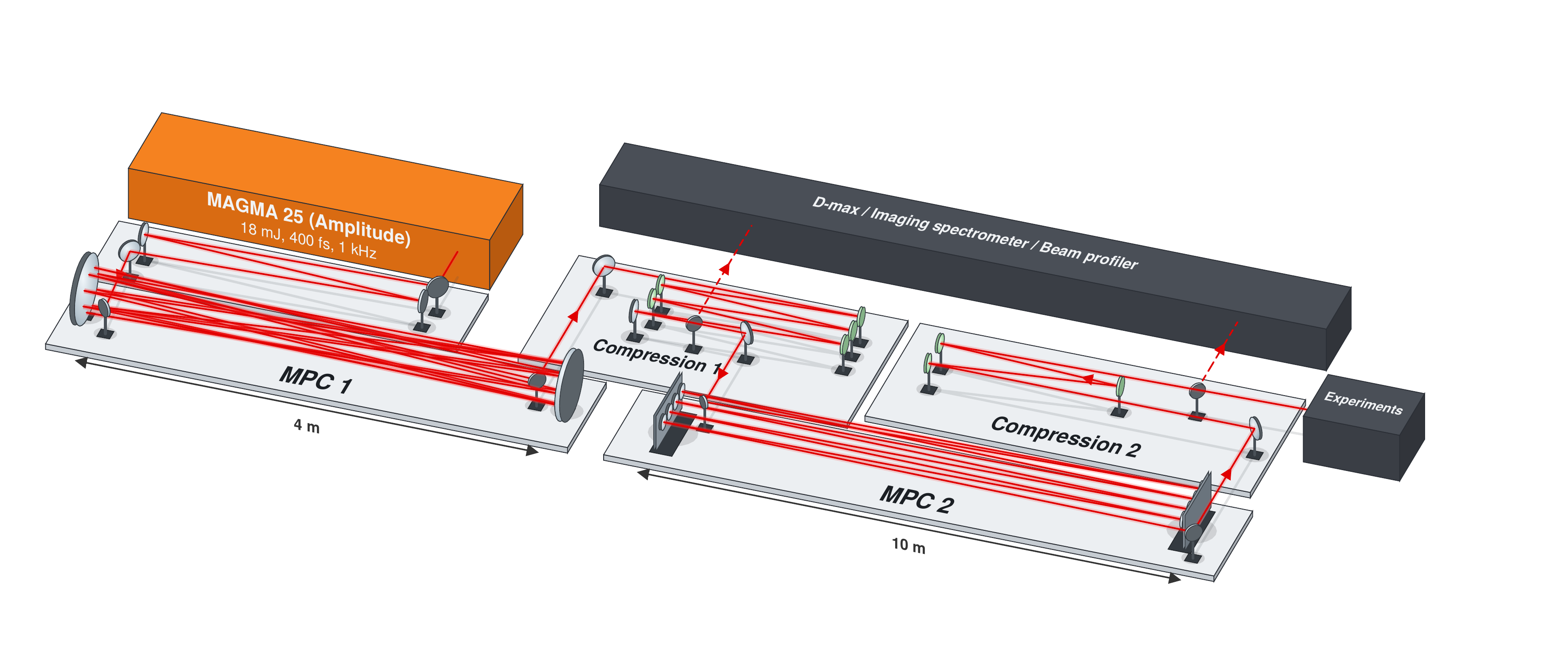}
    \caption{Scheme of the experimental setup. MPC 1 is filled with 140 mbar of argon, while Compression 1 and MPC 2 chambers are connected and filled with 500 mbar of helium. Compression 2 chamber is under secondary vacuum at 5 $\times$ 10$^{-6}$ mbar. The collimation after both MPCs and the second mode-matching telescope are not shown for the sake of clarity.}
    \label{fig:exp_setup}
\end{figure*}

The front end Yb-based laser (MAGMA25, \textit{Amplitude}) delivers 18 mJ, 400 fs-long laser pulses at 1 kHz. To reach sub-2-cycle duration (6.9 fs for a central wavelength of 1030 nm) from a 400 fs input, the post-compression system needs to yield a compression factor (C) of at least 60 which corresponds, using equation $\text{C} = 0.59\text{B} + 1$ from \cite{Escoto2022Temporal}, to a B-integral (B) of 100\ rad. To relax the compression system and take advantage of the higher reflectivity of dielectric mirrors, we divided the MPC post-compression setup into two stages, as shown in Fig.~\ref{fig:exp_setup}. The laser pulses first transits through an adjustable telescope used for nonlinear mode-matching of the beam to the eigenmode of the first cell \cite{Hanna2017Nonlinear}. The first MPC (MPC1) geometry is laid out in a nearly concentric Herriott-type configuration \cite{Herriott1964Offaxis}, comprising two low-dispersion broadband dielectric end-mirrors centered at 1030 nm of 76.2 mm diameter and -2 m radius of curvature (RoC), resulting in a 3.85 m MPC length. This configuration was chosen to limit ionization at focus and prevent damage on the end-mirrors. The distance between the mirrors is set such that MPC1 is re-entrant after the 30\textsuperscript{th} pass, and that after each roundtrip, the beam is shifted by two spots of the beam pattern on the mirrors ($N = 30$, $k = N-2$ according to \cite{Viotti2022Multipass}). This geometry allows for a maximum number of 29 passes in a chamber filled with 140 mbar of argon. The pulses are compressed in the same chamber as the second MPC (MPC2) to reduce the overall footprint and the amount of transmissive optics (windows). Compression is done using 9 chirped mirrors (\textit{Ultrafast Innovations}), compensating for -1800 fs$^2$ of dispersion. MPC2 is set up in an array-type configuration, where each 25.4 mm mirror, with -5 m RoC, is only used once, thereby enabling the possibility of experimenting with different materials/substrates/coatings for each pass. Indeed, the first MPC2 mirror is similar to the dielectric ones used in MPC1, as its bandwidth is still larger than the input laser spectrum. The remaining passes are done on enhanced silver mirrors with lower damage threshold, but the mode-matched $\sim$ 10 m-long cell ensures the fluence on the mirrors remains below 50 mJ/cm$^2$ for each pass. The multi-mirror approach helps improve overall transmission as the silver mirror reflectivity is between 98.5\% and 99\%, while that of the dielectric reaches 99.95\%. Also, this configuration is free of re-entrant conditions and geometrical constraints from the mirror size, and can accept many more beam sizes than if constrained by discrete N and k parameters. All 9 passes are implemented under 500 mbar of helium, ensuring minimal dispersion and no ionization at the beam waist. Lastly, the pulses are collimated at $\sim$ 12 mm diameter and enter the final compression chamber through an AR-coated 1\ mm MgF$_2$ window, where the beam is expanded to $\sim$ 25 mm diameter through a reflective telescope (enhanced silver mirrors). The compression is performed at full energy under secondary vacuum (< 10$^{-5}$ mbar) using uncoated fused silica wedges (only ones available at that time) and one pair of complementary chirped mirrors (\textit{Ultrafast Innovations}) that compensate for -120 fs$^2$ of dispersion. 

\begin{figure}[h]
    \centering
    \includegraphics[width=\linewidth]{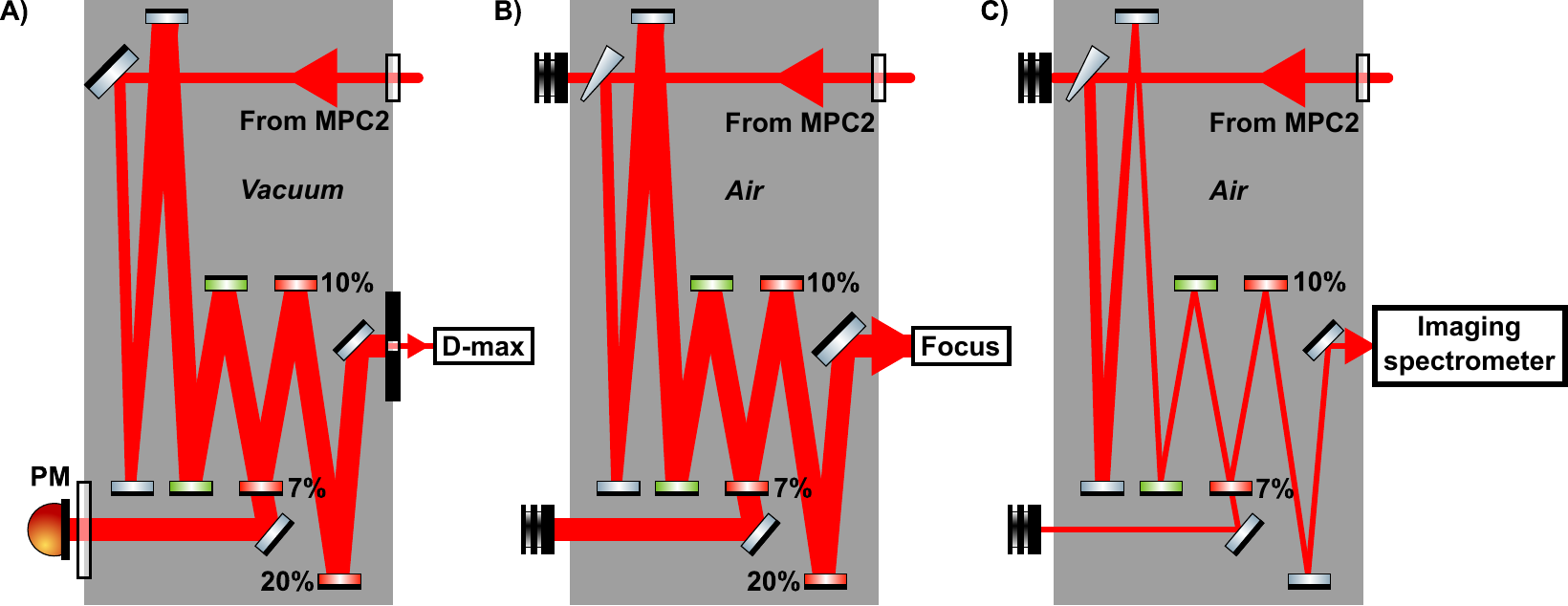}
    \caption{A) Setup of the compression chamber 2 for the measurement of the temporal profile. The beam is expanded, compressed and then sampled. The leak from the first beam splitter is used to assess the output energy. B) Setup for the measurement of the intensity profile at focus. The beam is sampled using an uncoated wedge, with the last mirror being the first one of setup A). C) Setup for the spatio-spectral homogeneity measurement. The beam is sampled with an uncoated wedge, reduced (by inverting the telescope), compressed and attenuated. Green mirrors: chirped mirrors. Red optics: ultra-broadband beam splitters, along with their reflectivity. PM: Powermeter.}
    \label{fig:setup-comp}
\end{figure}

Pulse diagnostics include a d-max (\textit{Sphere Ultrafast Photonics}) to characterize the temporal intensity profile, an imaging spectrometer (\textit{Femtoeasy}) to diagnose the spatio-spectral beam quality, and a beam profiler (\textit{Dataray}) to image the focused intensity profile. The final compression chamber (Compression 2) can be configured in 3 different ways to reliably perform each individual measurement, as shown in Fig.~\ref{fig:setup-comp}. In configuration A (Fig.~\ref{fig:setup-comp} A), the beam from MPC2 is expanded and temporally compressed at full energy under vacuum and then attenuated through a sequence of low-dispersion broadband beamsplitters (\textit{Laseroptik}). The portion of the full beam transmitted through the first beamsplitter is used for calibrated energy measurements of the compressed pulse outside the chamber. The attenuated beam is sent out of the chamber through an AR-coated, 1\ mm, MgF$_2$ window with 17 mm clear aperture, and the central portion of the beam exiting the chamber is then steered into the d-max device for temporal characterization in air. In configuration B (Fig.~\ref{fig:setup-comp} B), the exit window is removed and the first enhanced silver mirror after MPC2 is replaced by a wedge to further attenuate the pulse energy before compression in air. This attenuated full size beam is then used for focal intensity measurements in air outside the chamber. In configuration C (Fig.~\ref{fig:setup-comp} C), again in air, the telescope is inverted to downsize the 12 mm diameter beam from MPC2 to 6 mm in order to fit into the full aperture of the imaging spectrometer used outside the chamber for spatio-spectral homogeneity measurements. Here, the last beamsplitter is replaced by an enhanced silver mirror to achieve the required intensity level for the measurement. In configurations B and C, pulse compression does not affect the the corresponding measurements. 

\begin{figure}[h]
    \centering
    \includegraphics[width=\linewidth]{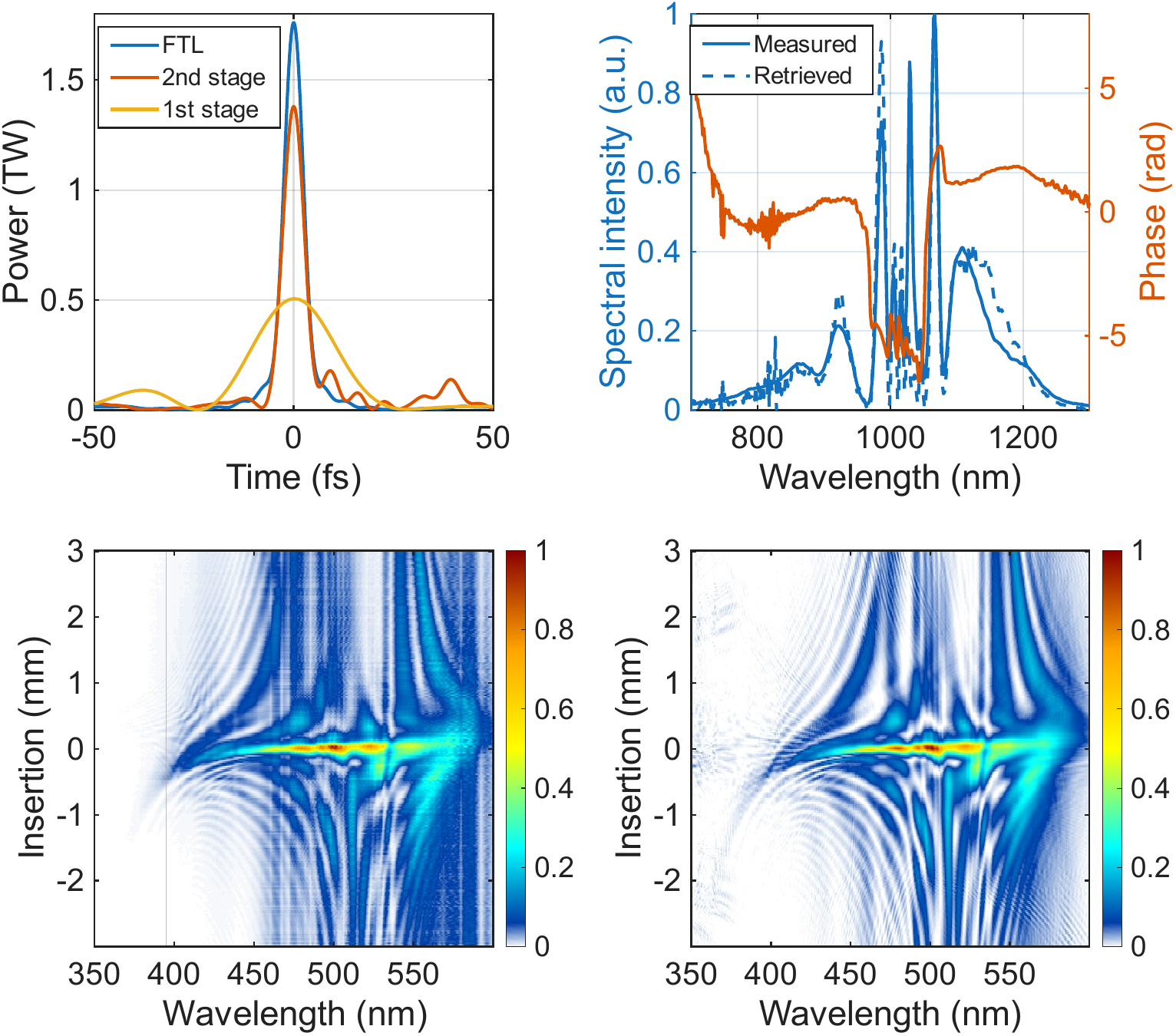}
    \caption{Temporal pulse measurements. Top left: Retrieved pulse and Fourier-Transform-Limited pulse after compression of MPC2 (d-max), along with the measured pulse after MPC1 (single-shot SHG-FROG). Top right: Measured and retrieved spectra, and retrieved spectral phase. Bottom left: Measured d-max trace. Bottom right: Retrieved d-max trace.}
    \label{fig:dmax_data}   
\end{figure}

The temporal profile reconstructed by the d-max measurement is shown in Fig.~\ref{fig:dmax_data}. The SHG signal had to be spectrally filtered to avoid leakage from the fundamental above 600 nm, due to slight tilting of the linear input polarization direction (close to p-pol). The use of a wedge to further attenuate the beam increased this tilt due to the difference in reflectivity between s-pol ($\sim$ 6\%) and p-pol ($\sim$ 0.8\%). Despite this underestimation of the spectral extension, the agreement between the measured and retrieved spectra is almost perfect, except for the central 1030 nm peak that is usually attributed to uncompensated higher-order spectral phase from the input laser. Finally, it is worth noticing that the central part of the laser spectrum appears out of phase with respect to the newly generated frequencies in MPC2. This corresponds to the output spectrum of the MPC1, and might cause visible degradation of the femtosecond level contrast. High-resolution temporal contrast measurements should give more insight into this effect. Despite this, the laser is compressed down to 6.2 fs (1.8 optical cycle at $\lambda_0 =1030$ nm) with an available energy of 12.5 mJ, yielding 1.4 TW peak power when normalizing the retrieved temporal profile. 

\begin{figure}[h]
    \centering
    \includegraphics[width=\linewidth]{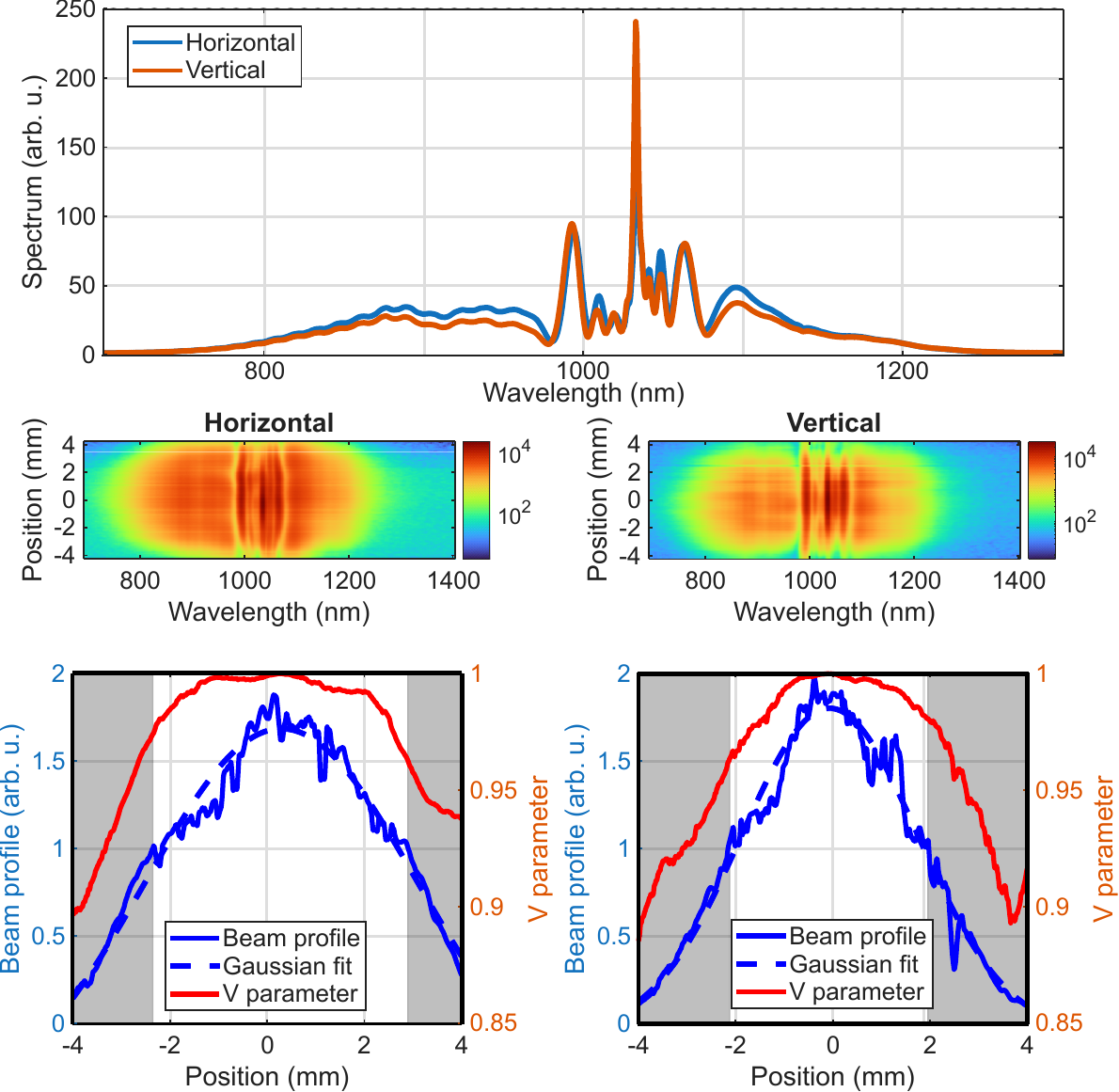}
    \caption{Spatio-spectral beam homogeneity measurements. Top: Integrated spectra for the horizontal and vertical axes. Center: Spectrograms from the horizontal (left) and vertical (right) axes, in logarithmic scale. Bottom: Corresponding integrated beam profiles (blue), along with a gaussian fit (dashed blue) and the corresponding computed V-parameter.}
    \label{fig:MISSdata}
\end{figure}

Fig.~\ref{fig:MISSdata} shows the results of the spatio-spectral homogeneity measurements performed on both horizontal and vertical beam axes. Here, the beam is downsized and focused onto the entrance slit of the imaging spectrometer using a cylindrical mirror. The integrated spectra look very similar, and agree with independent spectrometer measurements as well as with the d-max measurement. The computed V-parameter is clearly above 95\% at 50\% of the maximum intensity, and is above 90\% for both spectrograms.

\begin{figure}[h]
    \centering
    \includegraphics[width=\linewidth]{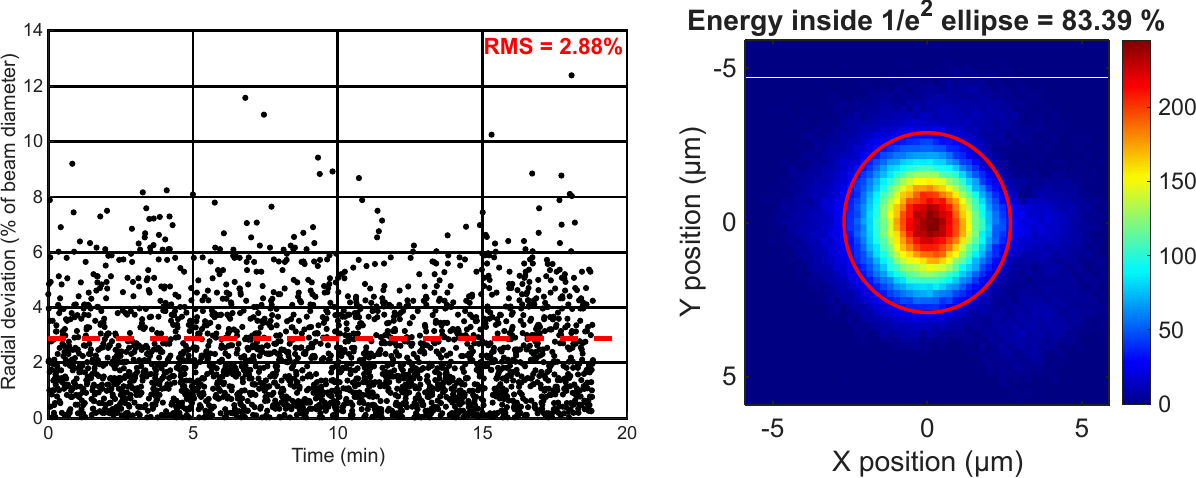}
    \caption{Left: RMS pointing stability, in percent of the beam diameter, recorded over 18 minutes. The red dashed line corresponds to an RMS value of 2.88\%. Right: 5.8 $\mu$m x 5.4 $\mu$m focal spot measured at the focus of a protected-silver coated 50.8 mm, f/2 90$^\circ$ off-axis parabola. The red ellipse shows the 1/e$^2$ intensity contour, that encircles 83.4\% of the total available pulse energy.}
    \label{fig:focus}
\end{figure}

The results of the focal intensity measurements are shown in Fig.~\ref{fig:focus}. The compressed attenuated full size beam is sent onto a protected-silver-coated 90-degrees off-axis parabola with 50.8 mm clear aperture and an effective focal length (EFL) of 10 cm. The beam focus is imaged onto a beam profiling camera (WinCamD, \textit{DataRay}) using a 10x microscope objective, and the imaging system is calibrated by inserting a diffraction mask into the beam prior to focusing. The focal spot displayed in Fig.~\ref{fig:focus}(right) shows no visible aberrations, the 1/e$^2$ intensity region encircles more than 83\% of the total energy and the focal spot is estimated to have a 1/e$^2$ diameter of 5.8 $\mu$m x 5.4 $\mu$m. Normalizing this focal spot size to the peak power from Fig.~\ref{fig:dmax_data} yields a peak intensity of 1.2 $\times$ 10$^{19}$ W/cm$^2$, and a normalized vector potential $a_0 = 0.85\sqrt{I[10^{18}W/cm^2]\lambda^2[\mu m]} = 3$, which is more than sufficient to perform LPA experiments at 1 kHz \cite{Faure:19}

Before each MPC, the laser is stabilized in position and pointing using a 4-axis combined motor/piezo beam stabilization system, to limit the couplings between beam movements, energy and broadening in both MPC stages (mainly caused by clipping on mirror edges). The pointing stability of the measured focal spot is shown in Fig.~\ref{fig:focus}(left). The RMS pointing fluctuation extracted from the raw data is of 0.18 $\mu$m, which corresponds to about 3\% of the beam diameter at focus. 

In conclusion, we have demonstrated a double-stage MPC post-compression system delivering 6.2 fs, 12.5 mJ pulses at 1 kHz repetition rate from a commercial Yb-based laser, reaching an effective peak power of 1.4 TW. To our knowledge, this constitutes the first demonstration of a terawatt-class, sub-2-cycle, kHz-repetition-rate post-compressed Yb laser system, in principle suitable for LPA experiments. Further compression towards the single-cycle limit will likely require alternative mirror designs supporting octave-spanning bandwidths or an additional MPC stage for enhanced spectral phase control. The maximum peak fluence allowed on the silver end-mirrors imposes a system footprint (total length $\sim$ 12 m) that is already at the limit of our available laboratory space. Scaling this approach in energy using the current setup could be achieved by using higher order Laguerre-Gaussian modes, as demonstrated in both gas \cite{Kaumanns2021Spectral, Ierano2026Energy} and bulk \cite{Koltalo2025Energy} MPC experiments, as they offer a more uniform fluence distribution than the fundamental Gaussian mode and relax peak fluence constraints without increasing cell size \cite{Hanna2025Higherorderand}. Scaling this approach to higher average powers by increasing the repetition rate will require the use of substrates with higher thermal conductivity \cite{Muller:21} and active cooling of the mirror substrates \cite{Seres2026Over} to preserve spatiotemporal pulse quality. Finally, high-dynamic range third-order autocorrelation measurements are currently underway to investigate potential issues concerning the temporal pulse contrast.

\begin{acknowledgments}
This project has received funding from the European Union’s Horizon 2020 Research and Innovation program under Grant Agreement 101004730, the Agence Nationale de la Recherche (ANR-24-CE92-0010 MILLSTREAMS) and from the Fondation de l'École polytechnique. F.S. and C.S. acknowledge the support of PACRI - Grant Agreement n. 101188004. The authors would like to thank the Institut Pierre Lamoure for continued support. The authors thank Sphere Ultrafast Photonics for the loan of the d-max device and for their help in interpreting the measurements.
\end{acknowledgments}

\bibliography{references.bib}
\end{document}